\documentclass[12pt]{article}
\usepackage{putex}
\usepackage{graphicx}
\usepackage{latexsym}

\newcommand{\beq}{\begin{equation}}
\newcommand{\eeq}{\end{equation}}
\newcommand{\beqa}{\begin{eqnarray}}
\newcommand{\eeqa}{\end{eqnarray}}

\newcommand{\lap}{\lower.5ex\hbox{$\; \buildrel < \over \sim \;$}}
\newcommand{\gap}{\lower.5ex\hbox{$\; \buildrel > \over \sim \;$}}

\begin{document}

\preprint{hep-th/0407097 \\ PUPT-2123}

\institution{PU}{Joseph Henry Laboratories, Princeton University, Princeton, NJ 08544}

\title{Cosmology with a dynamically screened \\ scalar interaction in the dark sector}

\authors{Steven S. Gubser\footnote{e-mail: \tt ssgubser@Princeton.EDU} and P.J.E. Peebles\footnote{e-mail: \tt pjep@Princeton.EDU}}

\abstract{Motivated in part by string theory, we consider a modification of the $\Lambda$CDM cosmological model in which the dark matter has a long-range scalar force screened by light particles.  Scalar forces can have interesting effects on structure formation: the main example presented here is the expulsion of dark matter halos from low density regions, or voids, in the galaxy distribution.}


\maketitle

\section{Introduction}
\label{INTRODUCTION}

What if there were a long-range force other than gravity acting only on the dark matter? The idea of a long-range force from the exchange of a massless scalar has a long history, with more recent attention on a scalar force in the dark sector that avoids the constraint from the E\"otv\'os experiment  (as reviewed in \cite{fp}). A screening length on a dark sector force generated dynamically by the presence of light particles with a Yukawa coupling to the scalar was discussed in \cite{fp}, and the picture of two dark matter species with a scalar force of attraction among like particles and repulsion among unlike particles was discussed in \cite{gpOne}. Here we combine the two notions, in the dark matter  lagrangian 
 \eqn{Particleaction}{
  {\cal L} &= {1 \over 2} (\partial\phi)^2 + 
   \bar\Psi_s i\slashed\nabla \Psi_s + \bar\Psi_+ i\slashed\nabla \Psi_+ + 
   \bar\Psi_- i\slashed\nabla \Psi_-  \cr &\qquad{}- 
   y_s\phi \bar\Psi_s \Psi_s- (m_+ + y_+\phi ) \bar\Psi_+ \Psi_-
    - (m_- - y_-\phi )\bar\Psi_-\Psi_-\,.
 }
The constants $m_\pm$ and $y_\pm$ are all positive.  The fermions $\Psi_\pm$ are the non-relativistic dark matter, and the quanta of the additional species $\Psi_s$ will be termed screening particles.  The screening particles are massless at $\phi=0$, and they can provide dynamical screening by pulling the field close to $\phi = 0$.  An advantage of this mechanism is that the screening length increases with the size of the universe, so that its length in co-moving coordinates is roughly constant: thus it can play  an important role in the formation and evolution of the galaxies at low redshift but not substantially affect the formation of the observed anisotropy of the 3~K thermal background radiation (the CMB) at decoupling at redshift $z\sim 1000$.  

We have two purposes in this continuation of our previous work \cite{fp,gpOne}.  First, we continue the exploration of possible astrophysical implications of the scalar-mediated force that emerges from the action \Particleaction, in this paper with particular attention to the properties  of voids. We argue that the repulsive force between dark matter particles with opposite scalar charge can make the voids more empty than the prediction of the $\Lambda$CDM model, and an apparently better approximation to what is observed \cite{voids}.  Second, we argue that the action~\Particleaction\ fits rather neatly into the framework of supersymmetry and string theory, although supersymmetry breaking as it is usually understood causes some problems.  In summary, we argue that a scalar-mediated, dynamically screened interaction in the dark sector has both theoretical motivation and possibly interesting observational consequences.

In the modified CDM model in~\cite{gpOne}, the dark matter could be described in terms of a scalar field coupled to two species of fermions, with an interaction lagrangian of the form
\eqn{Oldparticleaction}{                                                                                                                                                                                                                                                                                                                                                                                                                                                                                                                      
{\cal L}_{\rm int} = - m_0 e^{\beta \phi /M_{\rm pl} }\bar\Psi_1\Psi_1- m_0 e^{-\beta \phi /M_{\rm pl}} \bar\Psi_2\Psi_2\,,
 }
where $m_0$ is close to the Planck scale and $\beta$ is of order unity.  The scalar interacts only with these particles, giving rise to an inverse square law proportional to the product of ``scalar charges'' $Q_i = dm_i/d\phi$.  The total scalar charge density must vanish once $\phi$ relaxes to its equilibrium point, determined by an effective potential that includes contributions from the dark matter.  Additional species with couplings to the scalar do not change the picture much, provided they are heavy.

\begin{figure}
  \centerline{\includegraphics[width=3in]{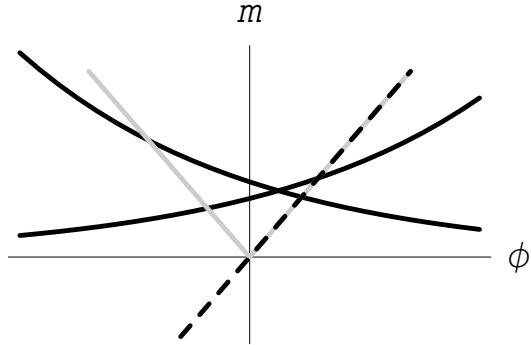}}
  \caption{The solid lines depict the masses for non-relativistic dark matter species included in the model of \cite{gpOne}.  The dashed line is $m(\phi)$ for the additional, light species $\Psi$, while the gray line shows the physical mass $|m(\phi)|$.  Including all three species leads to a model like \Particleaction, provided $\phi$ stays close to the zero of $m(\phi)$.}\label{figA}
 \end{figure}

The variant \Particleaction\ which we explore in this paper is based on the idea that the mass parameter for some species could pass through zero.  For example, the mass term for $\Psi_s$ is $-y_s\phi \bar\Psi_s \Psi_s$, but the overall sign of this term can be reversed by performing a chiral rotation, $\Psi_s \to e^{i\pi\gamma_5} \Psi_s$.  Thus for \Particleaction, the mass of $\Psi_s$ is $|y_s\phi|$.  The situation is summarized in figure~\ref{figA}.  If there are many $\Psi_s$ quanta compared to the number of $\Psi_\pm$ quanta, then $\phi$ is pinned near zero by the contribution of $\Psi_s$ to the effective potential.  The total scalar charge for $\Psi_\pm$ quanta need not be zero because the screening particles can make up the difference --- a first hint that $\Psi_s$ effectively screens the scalar interaction.

The organization of the remainder of the paper is as follows.  In section~\ref{STRINGS} we describe some reasons to regard scalar forces in the dark sector as well-motivated in string theory, and we construct supersymmetric generalizations and extensions of the action \Particleaction from ingredients that arise in compactifications of string theory.\footnote{The reader wishing to pass lightly over the string theory may find the summary in subsection~\ref{SUMMARY} helpful.}  In section~\ref{SCREENING} we briefly recapitulate the dynamical screening mechanism.  Aspects of the growth of structure in the dark matter distribution, in linear perturbation theory and a spherical model, are discussed in section~\ref{GROWTH}, with particular attention to the case where the two dark matter species have very different scalar charge-to-mass ratios. In section~\ref{VOIDS} we consider the possible application to the properties of voids in the distribution of normal galaxies.

\section{Supersymmetric and string theoretic models}
\label{STRINGS}

There are both top-down and bottom-up motivations to consider a link between string theory and scalar forces in the dark sector.

A top-down motivation is that there are moduli in string theory, and there are heavy objects whose mass depends on them, offering the opportunity to stabilize moduli not through vacuum effects but through the presence of some density of the heavy objects \cite{kp,bv}.  It is hard to see how this could work in the visible sector because of the tight constraints on fifth forces and on the variations of fundamental constants such as the fine-structure constant.  But in the dark sector, it is not only possible --- because of the considerably weaker constraints on the Equivalence Principle for dark matter particles --- but interesting for structure formation, as has been argued in \cite{fp,gpOne} and will be further argued in section~\ref{VOIDS}.

A bottom-up motivation is that, to have interesting effects, scalars in the dark sector must be very light, and the only obvious means to accomplish this is supersymmetry.  (Below we will review the situation when supersymmetry is broken).  More specifically, with the interactions in~\Particleaction\ present, it is unnatural for the potential $V(\phi)$ to be zero: absent some symmetry argument, the natural scale for $V(\phi)$ is the Planck scale.  So let us provisionally think of $\phi$ as a flat direction whose existence is protected by standard non-renormalization theorems of unbroken supersymmetry.\footnote{Another way to get a light scalar is to have it be a Goldstone boson or axion, but the Yukawa-type couplings in \Particleaction\ are forbidden.}  Actually, in a supersymmetric theory, massless scalars are complex, and there could be several of them.  But the physical consequences are fairly similar to the simplest case of one real scalar.

A further qualitative feature that \Particleaction\ provides is degrees of freedom (the screening particles) which become massless at a special value of the massless scalar $\phi$.  The spin and statistics of these particles is not important: the density in single particle phase space can be negligibly small, and the dark matter Compton wavelengths are very much smaller than length scales of interest in astronomy.  But the occurrence of additional massless particles at special points in the moduli space is a hallmark of string compactifications, and there are even some dynamical reasons to think that vacua with such particles present are preferred \cite{KLLMMS}.  Such vacua are often referred to as having enhanced symmetry because a particularly common example is to have some non-abelian gauge symmetry become unbroken.  In fact, the logic of \cite{KLLMMS} is that if the universe evolves through a configuration with extra massless particles, then these particles will be produced by quantum effects and will tend to prevent the further evolution from moving away from the point of enhanced symmetry.  One of our desiderata, to be described in more detail in section~\ref{SCREENING}, is that the screening particles should be large in number compared to the massive dark matter particles, and the mechanism of \cite{KLLMMS} will tend to make this so.\footnote{However, as we will describe in section~\ref{BOUNDS}, the screening particles need to be highly energetic---more so than the mechanism of \cite{KLLMMS} would seem likely to predict.} 

The massive dark matter particles need to have a ratio of scalar charge to mass on the order of $1/M_{\rm Pl}$, though with screening, it is interesting to consider such ratios an order of magnitude higher.  String moduli are associated with the gravitational physics of the extra dimensions, and the massive dark matter particles can be strings or branes stretched around some cycle or between some pair of other branes: then a charge-to-mass ratio on the order of $1/M_{\rm Pl}$ is indeed the natural range. 

In summary, there is an interesting convergence of lines of thinking toward models of the type \Particleaction. In the following subsections, we will consider two  supersymmetric models in more detail, note some pitfalls, and present one further argument in favor of light scalars even after supersymmetry breaking.  The first model, based on chiral superfields, fails to reproduce the physics of \Particleaction, but for interesting reasons which will guide us to a better model, presented in section~\ref{GAUGED}.

\subsection{A chiral model and the pitfall of D-terms}
\label{CHIRAL}

One choice for the dynamics is to replace $\phi$ and $\Psi_s$ with three chiral superfields, $\Phi_1$, $\Phi_2$, and $\Phi_3$, with canonical kinetic terms and a superpotential
 \eqn{TotalW}{
  W = y_s \Phi_1 \Phi_2 \Phi_3 \,.
 }
Assuming standard kinetic terms, the lagrangian for the component fields of the $\Phi_i$ only is
 \eqn{LSUSY}{
  {\cal L} &= \sum_k \left( |\partial\phi_k|^2 + 
   {1 \over 2} \bar\Psi_k i \slashed\partial \Psi_k \right) -
   \left( y_s\phi_1\psi_2\psi_3 + y_s\phi_2\psi_3\psi_1 +
      y_s\phi_3\psi_1\psi_2 + h.c. \right) \cr&\qquad{} - 
    y_s^2 |\phi_1\phi_2|^2 - y_s^2 |\phi_2\phi_3|^2 - 
      y_s^2 |\phi_3\phi_1|^2 \,,
 }
where we decompose a Majorana spinor as $\Psi = \left( \psi_\alpha \atop \bar\psi^{\dot\alpha} \right)$, and $\psi_1\psi_2$ means $\psi_1^\alpha \psi_{2\alpha}$.  In the flat directions for the theory \LSUSY\  one of the $\phi_i$ is non-zero and the others vanish.  Suppose $\phi_1 \neq 0$.  Then the bosonic and fermionic fields of $\Phi_2$ and $\Phi_3$ acquire masses $|y_s\phi_1|$, but $\phi_1$ is protected against acquiring a potential by the supersymmetry.  Thus $\phi_1$ plays the role of $\phi$ in the lagrangian \Particleaction, and the component fields of $\Phi_2$ and $\Phi_3$ play the role of $\Psi_s$.

We may reintroduce the non-relativistic dark matter species $\Psi_\pm$ as components of two additional chiral superfields $\Phi_\pm$, again with standard kinetic terms, and with additional terms in the superpotential of the form
 \eqn{AdditionalW}{
  \delta W = {1 \over 2} (m_+ + y_+ \Phi_1) \Phi_+^2 + 
   {1 \over 2} (m_- - y_- \Phi_1) \Phi_-^2 \,.
 }
Instead of writing out the component lagrangian in detail, let us simply remark that $\phi_\pm$ and $\Psi_\pm$ wind up having masses $m_\pm \pm y_\pm \phi_1$, so they play the role of the fields $\Psi_\pm$ in~\Particleaction: their quanta are the non-relativistic dark matter.

\begin{figure}
  \centerline{\includegraphics[width=4in]{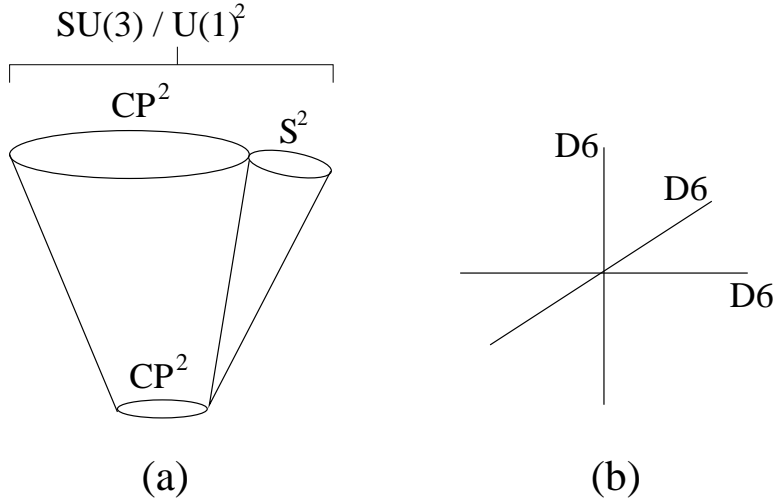}}
  \caption{a) A cartoon of the $G_2$ manifold asymptotic to a cone over $SU(3) / U(1)^2$.  The topology of the manifold is ${\bf CP}^2 \times {\bf R}^3$.  b) M-theory dynamics on the $G_2$ manifold is equivalent to type IIA string theory with a three D6-branes intersecting at a point.}\label{figB}
 \end{figure}

In \cite{aw}, strong evidence was presented for the claim that M-theory on ${\bf R}^{3,1}$ times a particular seven-manifold of $G_2$ holonomy, asymptotic to a cone over $SU(3) / U(1)^2$, has for its low-energy dynamics the supersymmetric model described by three chiral superfields with superpotential $y_s \Phi_1 \Phi_2 \Phi_3$ (see figure~\ref{figB}a).  The $G_2$ cone is not compact: it has infinite volume.  So gravity is decoupled and the theory has only global supersymmetry.  But asymptotically conical $G_2$ manifolds describe the local geometry of compact $G_2$ manifolds near isolated pointlike singularities, and in the compact case gravity ceases to decouple and supersymmetry becomes local.  Thus it would seem that a $G_2$ compactification that includes a singularity where some $SU(3) / U(1)^2$ shrinks would be a good candidate for an embedding of \Particleaction\ in M-theory.  The massive dark matter particles could be wrapped M2-branes that pass through or close to the singularity.  The appeal of this picture is enhanced by work (see for example \cite{csuTalk,AchSchool}) indicating that $G_2$ manifolds with singularities (more precisely, intersecting loci of ADE singularities) offer a way of constructing quasi-realistic theories that include the Standard Model.

However, there is a significant hitch.  Consider the type~IIA description of the cone over $SU(3) / U(1)^2$: three D6-branes intersecting at a point (see figure~\ref{figB}b).  The fields $\Phi_1$, $\Phi_2$, and $\Phi_3$ are strings stretching between pairs of branes.  After compactification, the gauge fields on the D6-branes must be expected to become dynamical.  Their coupling to the fields $\Phi_i$, together with ${\cal N}=1$ supersymmetry, implies a D-term potential of the form
 \eqn{Dterm}{
  V_D = g_1^2 \left( |\phi_2|^2 - |\phi_3|^2 \right)^2 + 
   g_2^2 \left( |\phi_3|^2 - |\phi_1|^2 \right)^2 + 
   g_3^2 \left( |\phi_1|^2 - |\phi_2|^2 \right)^2 \,,
 }
where the $g_i$ are the gauge couplings for each D6-brane.  The potential \Dterm, together with the F-term potential exhibited in the second line of \LSUSY, leaves no flat directions: for example, if one sets $\phi_2=\phi_3=0$, then the potential in the $\phi_1$ direction is $(g_1^2 + g_2^2) |\phi_1|^4$.  The fact that D-term potentials can lead to directions where a quartic self-interaction (with coupling of order unity) keeps the field pinned to zero is interesting for the version of chameleon scalar fields described in \cite{gKhoury}, but disqualifies the construction as something that could change large-scale structure formation.  (Note, for instance, that even a very low concentration of dark matter particles of a particular scalar charge would screen the scalar force to uninterestingly small length scales via the chameleon mechanism \cite{KhouryWeltmanOne}).

Before the D-term potential \Dterm\ spoils the story, what makes the model \LSUSY\ work as a model that includes screening particles is that the space of supersymmetric vacua is not a manifold: three branches join at the point where $\phi_1=\phi_2=\phi_3=0$.  Many supersymmetric field theories exhibit a similar feature of multi-branched moduli spaces, and a number arise from string theory constructions.  We will return to examples in this general class in section~\ref{GAUGED}.

\subsection{Supersymmetry breaking}
\label{NONSUSY} 

If supersymmetry is broken, then presumably a potential for otherwise massless scalars is generated at the soft breaking scale.  Following conventional notions of supersymmetry breaking, one would predict that this scale could be as low as about $0.1\,{\rm eV}$ --- some $28$ orders of magnitude below the Planck scale.  But for the scalar force to have a range on the order of a megaparsec, we need the scalar mass to be some $28$ orders of magnitude smaller still!  Thus, in considering supersymmetric theories with flat directions as candidate ingredients in a dark sector with scalar forces, we are entertaining the notion that supersymmetry is unbroken in the dark sector, or at least that the scalar somehow remains very light.  This seems to be in clear conflict with standard notions of supersymmetry breaking, which hinge in part on the concept of field theoretic naturalness.

On one hand, in light of the cosmological constant problem, we are disinclined to rule out a construction solely on the basis of difficulties that arise after supersymmetry is broken --- particularly when it provides an interesting class of alternatives to $\Lambda$CDM.

On the other hand, it is intriguing to note that in AdS/CFT \cite{juanAdS,gkPol,witHolOne}, scalar fields whose masses are comparable to the scale of spacetime curvature are dual to operators in the field theory whose dimension is order unity.  The precise relation for $AdS_5/CFT_4$ is $m^2 L^2 = \Delta(\Delta-4)$ \cite{gkPol,witHolOne}, where $L$ is the radius of curvature, i.e.~the analog of the Hubble length in the static $AdS_5$ geometry.  AdS/CFT is a general non-perturbative equivalence whose validity does not rely on supersymmetry.  Were it not for this duality, the presence of scalar fields with mass $\sim 1/L$ (even after quantum corrections) would be very surprising in a limit where $L$ is much larger than the Planck scale.  With the duality in hand, such scalar fields are precisely as natural as scalar operators whose dimension remains finite in the large $N$ limit that corresponds to $L \gg 1/M_{\rm Pl}$.

It has been suggested \cite{AndydS} that AdS/CFT may be extended to a correspondence between $dS_4$ and a three-dimensional Euclidean quantum field theory.  In such a correspondence, scalar operators in the field theory with finite dimension would correspond precisely to scalars whose masses are on the order of the Hubble scale --- the main ingredient that we need for our story.\footnote{This line of argument, starting with scalars in AdS, was suggested to us by C.~Vafa, based on work with collaborators \cite{VafaUnpublished}.  We thank him for his permission to review it here.}  Eventually, this welcome conclusion would have to be reconciled with the loop calculations that support the field-theoretic notions of naturalness.  For now, we take it as a good reason to believe that light scalars are indeed possible in a quantum theory of gravity.\footnote{The loophole that permits light axions would imply rather special forms for the higher-point correlators of the dual operators.  The above argument could be more precisely phrased as saying that light scalars with non-derivative couplings are dual to a class of operators whose dimensions should not all become large in a large $N$ limit.}

\subsection{The Kahler potential}
\label{KAHLER}

Even before supersymmetry breaking, it is possible to run into trouble with coupling to the visible sector.  The difficulty is the Kahler potential $K$, which can receive quantum corrections at any order in perturbation theory.  

Let us recall why it is important to have some control over the Kahler potential \cite{gpOne}.  Yukawa couplings in the component lagrangian are accompanied by a factor $e^{K/2}$, so quark masses have some dependence on scalars in the dark sector.  The proton mass will also depend on such scalars: if 
 \eqn{Kgeneral}{
  K = \hbox{(constant)} + \Re(\xi\phi)/M_{Pl} -
   |\phi|^2/M_{Pl}^2 + \ldots \,,
 }
where $\xi$ is a dimensionless complex number, then
 \eqn{ProtonMass}{
  m_p \approx \bar{m}_p + \epsilon_p 
   \left[ {\Re\xi\phi \over 2M_{Pl}} - 
    {|\phi|^2 \over M_{Pl}^2} + 
    {(\Re\xi\phi)^2 \over 8M_{Pl}^2} \right] \,,
 }
where $\bar{m}_p$ is the proton mass when $\phi = 0$, and $\epsilon_p$ is of order the $u$ and $d$ quark masses, that is, approximately $7\,{\rm MeV}$.  In order not to violate current experimental constraints on violations of the Equivalence Principle, it is necessary to have both $|\phi/M_{Pl}| \lsim 2 \times 10^{-5}$ and $|\xi| \lsim 2 \times 10^{-5}$.  The first is not a problem based on estimates presented in \cite{gpOne}: the  deviation of $\phi$ from its background value is plausibly less than $10^{-6} M_{\rm Pl}$.  The second appears to be more difficult to ensure: it amounts to the statement that just when extra massless modes appear, there is no linear term in the Kahler potential.  If $\phi$ is charged under a gauge symmetry, then this linear term is impossible --- but then one has the D-term potential associated with the gauge coupling to contend with.  It is perhaps illustrative to consider the case of the cone over $SU(3) / U(1)^2$.  There, at the classical level, the Kahler potential is $K=-3 \log V$ where $V$ is the volume of the $G_2$ manifold and $\phi$ is one of the scalars $\phi_i$ in a regime where the singularity is resolved enough to avoid Planckian curvatures.  The cone volume is maximized when it is singular, leading us to expect $K \sim \hbox{constant} - \Phi^\dagger \Phi$: this is confirmed by the absence of the linear term because of gauge symmetries.

\subsection{Models with a non-abelian gauge interaction}
\label{GAUGED}

Let us now describe supersymmetric field theories generalizing \Particleaction\ which avoid the pitfalls of D-terms and linear terms in the Kahler potential described in previous subsections.

The combination of D-term and F-term potentials can leave protected flat directions in the space of supersymmetric vacua.  As an example, consider $SU(2)$ gauge theory with one adjoint flavor $X^a{}_b$ and several fundamental and anti-fundamental flavors, $q_i^a$ and $\tilde{q}_{ia}$.  The renormalizable superpotential\footnote{The superpotential \eno{Wseveral} is not the most general one allowed by renormalizability, but further generalizations of it are not particularly interesting for our purposes.} is
 \eqn{Wseveral}{
  W = y_s \tilde{q}_{ia} X^a{}_b q_i^b + 
   {1 \over 2} M_{ij} \tilde{q}_{ia} q_i^a + 
   {1 \over 2} m_X X^a{}_b X^b{}_a \,.
 }
If $M_{ij} = 0 = m_X$, then the F-term potential vanishes either when $X=0$ and $\tilde{q}_{ia} q_i^b = 0$ (referred to as the Higgs branch because $\tilde{q}$ and $q$ are allowed to be non-zero, which completely breaks the gauge symmetry), or when $q = 0 = \tilde{q}$ (referred to as the Coulomb branch because non-zero $X$ breaks the gauge group to $U(1)$).  The D-term potential adds quartic self-interactions on both branches, but flat directions remain: in particular, $V_D = \tr [X,X^\dagger]^2$ on the Coulomb branch, which (after accounting for gauge freedom) leaves a single-complex-dimensional moduli space.  

The dynamics of D3-branes near D7-branes is described by a field theory similar to \Wseveral: see for example \cite{PolVolTwo} for details.  It has ${\cal N}=2$ supersymmetry and moduli space composed of a Higgs and Coulomb branch that meet at the point where gauge symmetry is restored.

Provided $m_X=0$ and $M_{ij}$ has zero eigenvalues, theories like \Wseveral\ are good candidates for supersymmetric extensions of \Particleaction.  The scalar $\phi$ is replaced by the Coulomb branch, and the screening particles are both the $q_i^a$ and $\tilde{q}_{ia}$ and the non-abelian gauge bosons that become massless when $X=0$.  These particles interact with each other, but this does not substantially change the dynamical screening phenomenon which we review in section~\ref{SCREENING}.  Furthermore, gauge invariance prevents linear terms in $X$ in the Kahler potential.  And if $M_{ij}$ has some eigenvalues near the Planck scale, the corresponding $q_i^a$ and $\tilde{q}_{ia}$ can play the role of massive dark matter.

The above analysis is substantially unaffected by quantum loops provided the gauge interaction does not become strong in the infrared.  This is guaranteed if there are enough light fundamental flavors $q_i$ and $\tilde{q}_i$. 

The theory \Wseveral\ is evocative of ${\cal N}=2$ supersymmetric theories: with special choices of the parameters in \Wseveral\ --- in particular, $m_X=0$ --- it becomes ${\cal N}=2$ $SU(2)$ gauge theory with several hypermultiplets in the fundamental representation.  In the spirit of minimalism, it is interesting to inquire whether we can make do with the simplest such gauge theory, namely the case where there are no $q$ and $\tilde{q}$ fields.  This theory has strong coupling in the infrared, but the dynamics is well understood due to \cite{sw}.  There are special points on the Coulomb branch where dyons become massless, but the $SU(2)$ gauge symmetry is never restored (unlike the case when there are enough matter fields to make the gauge theory free in the infrared).  The $U(1)$ gauge boson is in a vector multiplet with a scalar $a$, where roughly $X = a J_3$ with $J_3$ a generator of $SU(2)$.  At the point where magnetic monopoles become massless, one may dualize to a magnetic description in which the light degrees of freedom are (in ${\cal N}=1$ language) $A_D$, $M$, and $\tilde{M}$ with a superpotential $W \propto \tilde{M} A_D M$.  It looks promising to regard $a_D$ as the scalar $\phi$ in \Particleaction\ and to let $M$ and $\tilde{M}$ be the screening particles.  Other dyons might be the massive dark matter.  But the Kahler potential --- which is constrained by ${\cal N}=2$ supersymmetry --- has a linear term in $A_D$, and this spoils the story as we have seen in section~\ref{KAHLER}.\footnote{The offending term in the Kahler potential arises because there is a linear term in the prepotential: $a = {\partial {\cal F}_D / \partial a_D} \sim a_0 + {i \over \pi} a_D \log a_D$ with nonzero $a_0$, and the Kahler potential is $\Im {\partial {\cal F}_D/\partial A_D} A_D^\dagger$.}

Finally, theories with ${\cal N}=4$ supersymmetry also have protected flat directions, extra massless particles at enhanced symmetry points of spins $0$, $1/2$, and $1$, and a protected Kahler potential.  Moreover, they are realized on coincident D-brane world-volumes (in the absence of fluxes or other supersymmetry breaking effects).  But explicit masses for the heavy dark matter particles would not respect the ${\cal N}=4$ supersymmetry.

\subsection{Summary}
\label{SUMMARY}

Let us summarize:
 \begin{itemize}
  \item If the dark sector is supersymmetric, massless scalars with a non-derivative coupling to non-relativistic dark matter become possible.
  \item Supersymmetry breaking in the dark sector at the lowest natural scale, roughly $0.1\,{\rm eV}$, tends to give scalars a mass on this order.  Thus, naturalness arguments pose a problem for long-range scalar forces.
  \item There are hints from AdS/CFT that scalars with mass comparable to the curvature scale are possible even without supersymmetry, despite the naturalness argument.
  \item Absent an experimentally testable understanding of supersymmetry breaking, we propose to search for viable models with supersymmetry, ultimately with a string theory origin.  Viable means that the model shouldn't have problems before supersymmetry breaking.
  \item One such problem is the lifting of flat directions by D-terms when a global symmetry becomes gauged: this is the fate of the $W = y_s \Phi_1 \Phi_2 \Phi_3$ model considered in section~\ref{CHIRAL}.
  \item Another such problem is the presence of linear terms in the Kahler potential at the point where extra massless particles arise: this is the fate of the model based on the Seiberg-Witten solution considered at the end of subsection~\ref{GAUGED}.
  \item Both problems are avoided by a class of models, considered in subsection~\ref{GAUGED}, with a cubic superpotential and an adjoint chiral superfield which provides the flat directions.
  \item Models of this general type describe the dynamics of D3-branes near D7-branes, so there is good reason to expect that they arise in quasi-realistic compactifications of string theory.
 \end{itemize}
One may reasonably ask whether supersymmetry and/or string theory are necessary to the discussion.  Indeed, one may take a completely phenomenological approach; yet it seems that string theory provides just the right ingredients for dynamically screened scalar forces acting with a strength comparable to gravity.  Those ingredients are: moduli, supersymmetry, points of enhanced symmetry, and stable heavy objects such as wrapped branes.

The bottom line is that string theory can help inform studies of an interesting class of generalizations of $\Lambda$CDM.  The issue of supersymmetry breaking is not lightly to be dismissed.  But we feel justified in bending the rules when the output is a class of models which may be tested against astrophysical data.

\section{Dynamical screening}
\label{SCREENING}

We turn now from the analyses in section~\ref{STRINGS} of the theoretical basis for the lagrangian \Particleaction\ to a review and extension of the application of the lagrangian to the dynamical screening mechanism developed in \cite{fp}.  In subsection~\ref{BASIC} we recapitulate the basic mechanism.  In subsection~\ref{EXCLUDE} we show that the screening particles can be excluded from halos formed from dark matter with a scalar charge.  In subsection~\ref{BOUNDS} we consider bounds on the various parameters of the model.

\subsection{The screening mechanism}
\label{BASIC}

We assume that the propagating (source-free) part of $\phi$ may be neglected, and that the number density $n_s$ of the screening particles (the quanta of $\Psi_s$) is large enough to drive $\phi$ close to zero almost everywhere, so the screening particles are relativistic. Since the source for $\phi$ --- in the inhomogeneous distribution of the massive dark matter --- varies with position on scales much larger than the Compton wavelength of the screening particles and much smaller than the Hubble length, the motion of a screening particle is close to adiabatic. That is, its energy $\epsilon$ is conserved, apart from the effect of the general expansion of the universe. The screening particle mass is $m_s = y_s |\phi|$ (assuming as usual that $y_s>0$), its velocity is $v$, and its conserved energy is $\epsilon = m_s/\sqrt{1-v^2}$.  

The equation of motion for $\phi$, neglecting source terms from the non-relativistic species, is
 \eqn{PhiEOM}{
  \Box\phi = y_s \bar\Psi_s\Psi_s
   = y_s n_s \langle\sqrt{1-v^2}\rangle\sgn\phi
   = {y_s^2 n_s \over \epsilon_s} \phi \equiv 
   {1 \over r_s^2} \phi \,.
 }
In the second equality of \PhiEOM, one observes that $\bar\Psi_s\Psi_s = n_s$ when $v=0$, and that both $\bar\Psi_s\Psi_s$ and $n_s \sqrt{1-v^2}$ are scalars under Lorentz transformations.  The next equality preserves the sign of $\phi$, because the screening particles have positive energies. The characteristic mean particle energy, from the average of the reciprocal Lorentz factor, is $\epsilon _s$. The last expression defines a screening length,
 \eqn{rsDef}{
  r_s \equiv \sqrt{\epsilon_s \over y_s^2 n_s} \,.
 }
Now, operating in the quasi-static approximation  --- where we drop time derivatives of $\phi$ --- let us re-introduce the two non-relativistic dark matter species, to get the wave equation
 \eqn{WithDarkMatter}{
  \nabla^2 \phi = y_+ n_+(t,\vec{r}) - y_- n_-(t,\vec{r}) + \phi/r_s^2 \,,
 }
where $n_\pm$ are number densities and $y_\pm$ are again positive. The first two terms on the right-hand side act as scalar charge densities and the last term produces the exponential cutoff in the scalar interaction of the dark matter.

The spatial mean of the wave equation \WithDarkMatter\ gives the relation
\eqn{meanwaveeq}{
\eta\equiv {-y_+\bar n_+ + y_-\bar n_-\over y_s\bar n_s} = 
 \langle\langle\sqrt{1-v^2}\rangle \, \sgn\phi\rangle\,,
}
where the inner brackets indicate averaging over the ensemble of screening particles to obtain a typical inverse Lorentz factor, while the outer brackets indicate averaging over space.  We are assuming the magnitude of the ratio $\eta$ of scalar charge densities is less than unity, so $\sqrt{1-v^2}$ may be small almost everywhere. This means the screening particles are allowed to move so most remain relativistic, in positions where $\phi$ is locked close to zero. Since the particles are conserved (apart from the small fraction captured in black holes) this condition is independent of redshift. One sees also that, as noted earlier, when the screenng particles are present the mean scalar charge density $y_+\bar n_+ - y_-\bar n_-$ need not vanish.

It may be important that when $\phi$ is close to zero nearly everywhere, so the screening particle nearly fill space, the general expansion of the universe causes $\epsilon _s$ to vary inversely as the scale factor $a(t)$. This means the screening length $r_s$ increases linearly with $a(t)$. Thus screening can occur at astronomically interesting length scales at low redshift but be insignificantly small at the epoch of formation of the observed anisotropy of the thermal cosmic microwave background radiation (the CBR).

Let us consider finally the evolution of the peculiar velocities of the screening particles. When the particles are free to roam the general expansion of the universe stretches de Broglie wavelengths in proportion to the expansion parameter, $a(t)$. Equation \meanwaveeq\ says the particle mass is $m_s=|\eta|\epsilon _s$, so the momentum of a particle varies as 
\eqn{momentumscaling}{
p = \sqrt{\epsilon ^2-m_s^2} = \epsilon\sqrt{1 - (\eta\epsilon_s/\epsilon)^2}
\propto a(t)^{-1}.
}
If the bulk of the screening particles are relativistic to begin with, then their momenta and energies both scale with the expansion of the universe as $a(t)^{-1}$, the mass scales as $a(t)$, and the screening particle peculiar velocities are independent of redshift.

\subsection{Excluding screening particles from halos}
\label{EXCLUDE}

When clustering in the spatial distribution of the scalar charge density has grown large enough, $\phi$ is pulled far enough from zero that $y_s|\phi| > \epsilon_s$. Where this happens the screening particles are excluded: their mass $y_s|\phi|$ in this region is greater than their characteristic energy.  Screening ceases in such a region. We find a condition for the screening particles to roam freely by using the solution to \WithDarkMatter\ valid in the limit where the length scale $r_v$ for variation of the source density $y_+n_+-y_-n_-$ is much larger than the screening length $r_s$. In this limit the left hand side of \WithDarkMatter\ may be neglected, and we have
 \eqn{SlowSources}{
  \phi \approx r_s^2 (-y_+n_+ + y_-n_-) \,.
 }
Then the condition $y_s|\phi| < \epsilon_s$ with the definition \rsDef of $r_s$ is
\eqn{conditon}{
|\eta | = {|y_+n_+(t,\vec{r}) - y_-n_-(t,\vec{r})| \over y_s n_s} < 1.
} 
In the opposite limit, where the scale $r_v$ of clustering of the dark matter is small compared to $r_s$, \WithDarkMatter\ becomes the Poisson equation, and the free roaming condition may be approximated as $|\eta | \lap (r_s/r_v)^2$, that is, it allows a larger density contrast. 

At high redshift, where the dark matter distribution is close to homogeneous, \conditon\ is satisfied everywhere if, as we are assuming, $|\eta |$ in \meanwaveeq\ is small. At low redshift, where the dark matter is strongly clustered, a region where $n_+$ is large will tend to drive out the $\Psi_-$ particles, so that $y_+n_+ \gg y_-n_-$.  Here we may write the condition that the screening particles roam freely in the presence of an overdensity of $\Psi_+$ particles with size $r_v\gg r_s$ as
 \eqn{condition1}{
  {y_sn_s\over y_+\bar n_+} > 
  {n_+(t,\vec{r}) \over \bar{n}_+}\equiv 1 + \delta_+(t,\vec{r}) 
   \,,
 }
where the number density $n_+(t,\vec{r})$ is expressed in terms of the mean density $\bar{n}_+(t)$ and the contrast $\delta_+(t,\vec{r})$. If $y_sn_s/y_+\bar n_+= 2$, for example, the screening particles may fill space in the early universe, when the mass density fluctuations are linear, and up to the time when the fluctuations start to become nonlinear on scales comparable to $r_s$. If $y_sn_s/y_+\bar n_+= 100$ the screening would last into significantly nonlinear mass fluctuations, but would be broken today within the nominal virial radius $r_v\sim 300$~kpc of a normal galaxy such as the Milky Way, unless $r_s\gg r_v$.

\subsection{Bounds on parameters}
\label{BOUNDS}
 
Another condition to consider is that the relativistic energy density $\epsilon _sn_s$ in the screening particles must not be large enough to spoil the standard model for the origin of prestellar helium and deuterium. We update the constraint in \cite{fp} under the assumption that the two non-relativistic dark matter components have the same mean densities of mass and scalar charge. 

The result of expressing $\rho_s/r_s^2$ in terms of the Hubble parameter $H_o$ and the energy fraction $\Omega_s$ in screening particles, with the assumption $n_+ m_+ = n_- m_- = \rho_m/2$, is 
 \eqn{hrs}{
  {\Omega_s \over (H_o r_s)^2} = {3 B_{++} \over 8}
   \left( {y_s n_s \over y_+ n_+} \right)^2 \Omega_m^2 
 }
where $\Omega _s$ and $\Omega _m$ are the density parameters in screening and the nonrelativistic $\Psi_\pm$ dark matter particles, and 
 \eqn{beta1}{
  {\rm B}_{++} = {y_+^2\over 4\pi Gm_+^2}
 }
is the ratio of the scalar and gravitational forces between two $\Psi_+$ particles that are much closer than the screening length.  The  constraint from light element nucleosynthesis in the early universe allows relativistic mass beyond the standard model equivalent to about one two-component neutrino family \cite{BBNS}, or $\rho_s \lap 0.23 aT_o^4$, where the present CBR temperature is $T_o=2.725$~K. At Hubble parameter $H_o=70$ km~s$^{-1}$~Mpc$^{-1}$ this condition is
\eqn{omegas}{
\Omega _s < 1.1\times 10^{-5}.
}
With dark matter density parameter $\Omega _m\simeq 0.3$ this bound in \hrs\ is
 \eqn{somecondition}{
  r_s<{80\over{\rm B}_{++}^{1/2}}{y_+n_+\over y_sn_s}\,\hbox{Mpc} \,.
 }
If ${\rm B}_{++}=1$, so the scalar force is as strong as gravity at separations $r\ll r_s$, and $y_+ n_+ =0.01 y_s n_s$, which would keep  the scalar pinned close to zero outside large galaxies, the screening length could be $r_s \sim 1\,{\rm Mpc}$, an interesting value for galaxy formation.

If the screening particles were the decay products of massive nonrelativistic particles with half life much larger than one minute then the present mean energy density in screening particles could be larger than the energy density in the CBR without upsetting light element nucleosynthesis. But if the half life were shorter than $10^5$~yr and $\rho _s$ were comparable to $aT_o^4$ it would increase the cosmological expansion rate at decoupling, pushing the peak of the CBR anisotropy spectrum to smaller scales, and upsetting the present apparent concordance of constraints on the cosmological parameters. The constraint \somecondition\ thus seems to be reasonably general. 

To develop a more complete set of upper and lower bounds on the various parameters of the model, let us note that the equalities $\rho_s = n_s \epsilon_s$ and $r_s^2 = \epsilon_s / y_s^2 n_s$ can be combined to give \hrs\ and a second relation,
 \eqn{TwoRelations}{
\qquad
  3 \Omega_s (H_o r_s)^2 = \left( {\epsilon_s \over y_s M_{\rm Pl}}
   \right)^2 \,.
 }
Let us start from a set of bounds intended to make the dark sector deviate in an acceptable but potentially interesting way from $\Lambda$CDM:
 \eqn{FirstBounds}{\seqalign{\span\TC}{
  {y_s n_s \over y_+ n_+} > 1, \qquad 
  30\, {\rm kpc} < r_s < 30\, {\rm Mpc}, \qquad
  {1 \over 2} < B_{++} < 10 \,.
 }}
The lower bound $y_s n_s/y_+ n_+ > 1$ ensures that if there is a scalar charge imbalance among the heavy particles over a very large region of the universe, the scalar field doesn't get pushed far away from zero.  We could tolerate a much smaller value of this ratio if $m_+(\phi)$ and $m_-(\phi)$ were perfectly linear, so that $n_+ m_+ + n_- m_-$ may be made independent of $\phi$: then an arbitrarily small $n_s$ locks $\phi$ close to zero.  But if departures from linearity in $m_\pm(\phi)$ arise from dimension $5$ operators like $(\phi^2 / M_{\rm Pl}) \bar \Psi_+ \Psi_+$, then we do need $y_s n_s/y_+ n_+ > 1$ to lock $\phi$.

With the cosmological parameters in \omegas\ and \somecondition , equations \hrs\ and \TwoRelations\ with \FirstBounds\ give \eqn{SecondBounds}{
  {y_s n_s \over y_+ n_+} < 4000, \qquad \Omega_s > 10^{-12},
  \qquad 1.3 \times 10^{-11} < {\epsilon_s \over y_s M_{\rm Pl}} < 
   9 \times 10^{-5} \,.
 }
A few comments are in order:
 \begin{itemize}
  \item The bounds \FirstBounds\ are deliberately inclusive, and it may not be interesting to simultaneously saturate the lower bound on $r_s$ and $B_{++}$, as we have done, for example, to derive both bounds in \SecondBounds.  Thus in particular, $y_s n_s/y_+ n_+$ is constrained to be in a fairly narrow window.
  \item As in \cite{gpOne}, we do not directly constrain $y_+$ or $m_+$, but rather their ratio, entering the above bounds through the ratio $B_{++}$ of scalar and gravitational forces.
  \item Likewise, we do not directly constrain $y_s$ or $\epsilon_s$, but rather their ratio.
 \end{itemize}

If one assumes $y_s \sim O(1)$, then the acceptable range \SecondBounds\ for $\epsilon_s$ represents large energies: so large that at a redshift of $10^{11}$ the typical particle's energy would be above the Planck mass.  One way to arrange this is to suppose that the screening particles arise from decays of very heavy particles which have long but finite lifetime: for masses on the order of $M_{\rm Pl}$, the lifetime should be on the order of a month.  An alternative --- 
making $\epsilon \sim 1\,{\rm eV}$ --- requires $n_s\sim 0.1$~cm$^{-3}$ and $y_s\sim 10^{-20}$, an exceedingly small value.

\section{The growth of structure}
\label{GROWTH}

We consider here the effect of the scalar interaction in two standard and simple models for structure formation: linear perturbation theory, which gives a good description of the growth of the clustering of mass and scalar charge at high redshift, and spherical symmetry, which illustrates effects of the nonlinear growth of structure. For the purpose of our considerations of the void phenomenon in the last subssection we will be particularly interested in the case where the two species of nonrelativistic dark matter have very different scalar charge-to-mass ratios. The dominant theme in this case is that the light component quickly responds to any scalar field gradient, and the massive component more slowly responds to the resulting distribution of the light component. We discuss the possible relevance to the void phenomenon in subsection~\ref{VOIDS}. This is not the only case of possible interest, of course. We briefly comment on other situations in section~\ref{DISCUSSION}. 

\subsection{Linear perturbation theory}
\label{LINEAR}

The mean mass density in non-relativistic species is
 \eqn{meanmassdensity}{
  \bar\rho_m(t) = m_+\bar n_+ + m_+\bar n_+ + \bar\rho_{\rm b} \,,
 }
where the last term represents the baryons.  Following standard practice we write $n = \bar{n}(1+\delta)$ for the spatially varying number density $n$ of each species and its density contrast $\delta$.  The total density contrast $\delta_m$ in non-relativistic species satisfies
 \eqn{masscontrast}{
  \bar\rho_m \delta_m = m_+\bar n_+\delta_+ + m_-\bar n_-\delta_- +
  \bar\rho_{\rm b}\delta_{\rm b} \,.
 }
In linear theory the condition $|\eta|<1$ in \conditon\ means $n_s$ is close to homogeneous if it began that way, so we can take $r_s$ to be independent of position. We can drop the time derivatives of $\phi$ because the inhomogeneities evolve slowly. Thus to determine $\phi$ we simply solve 
\WithDarkMatter.

Following \cite{gpOne}, it is straightforward to check that in linear perturbation theory the Fourier modes of the number density fluctuations evolve according to
 \eqn{FullPertEqn}{
  \ddot\delta_q + 2H \dot\delta_q = 4\pi G\rho _m  
   \sum_p \beta_{pq} f_p \delta_p \,.
 }
The dots represent derivatives with respect to proper time, so that Hubble's constant is $H = \dot{a}/a$.  The indices $q$ and $p$ run over the $\Psi_+$ particles, the $\Psi_-$ particles, and the baryons, and $f_p = n_p m_p / \sum_q n_q m_q$ is the mass fraction in species $p$.  The dimensionless scale-dependent quantities $\beta_{pq}$ are 
 \eqn{betaDef}{
  \beta_{pq} = 1 + {Q_p Q_q \over 4\pi G m_p m_q} 
   {k^2 \over k^2 + a^2/r_s^2} \equiv 1 + {\rm B}_{pq}(k) \,,
 }
where $\vec{k}$ is the wavenumber of the perturbation (so that $\vec{k}$ is dual to the coordinate $\vec{x}$ in the standard form of the cosmological line element, $ds^2 = -dt^2 + a(t)^2 d\vec{x}^2$), and the scalar charges are $Q_\pm = \pm y_\pm$ for the dark matter and $Q_b = 0$ for the baryons. Equation \FullPertEqn\ neglects radiation drag and the pressure of the baryon gas, both of which are good approximations on the scale of galaxies after recombination and prior to galaxy formation. We have remarked that $r_s$ grows proportionally with $a(t)$, that is, $a/r_s$ is constant, so $\beta_{pq}$ depends only on the species $p,q$ and the comoving wavenumber $k$.  The solutions are discussed in  \cite{gpOne}: one introduces the matrix $\Xi_{pq} \equiv \sqrt{f_p} \beta_{pq} \sqrt{f_q}$, which is real and symmetric, and whose eigenvalues $\xi_c$ and eigenvectors $c_q$ determine perturbations $\Delta_c = \sum_q c_q \sqrt{f_q} \delta_q$. When we can neglect the energy density in radiation and the cosmological  constant the modes vary with time as $\Delta_c \sim t^{2\gamma_{c\pm}/3}$, where $\gamma_{c\pm} = (-1\pm\sqrt{1+24\xi_c})/4$.

This analysis readily generalizes to more than two species of dark matter, each with its own scalar charge-to-mass ratio, and to more than one scalar field, by adding to the dimensions of the matrices ${\rm B}_{pq}(k)$ and $\Xi_{pq}$. The addition of gauge-mediated forces is easy, but, as discussed in \cite{gpOne}, it is not interesting because unlike gauge charges attract, as in an ordinary plasma, so charge separation is discouraged, just the opposite of the scalar interaction.

It is instructive to specialize to the case where baryons are a negligible fraction of the total matter density, the cosmological constant $\Lambda$ may be neglected, and there is overall charge neutrality in the non-relativistic species --- that is, $y_+ n_+ = y_- n_-$, or equivalently $f_+ y_+/m_+ = f_- y_-/m_-$.  Note that our assumptions imply $f_+ + f_- \approx 1$.  Then there is an adiabatic mode, where $\delta_+ = \delta_- = \delta_m$ and $\xi=1$, which grows as $\delta_m \sim t^{2/3}$ (the usual CDM result); and there is an isocurvature mode, where $\delta_{\rm iso} = \delta_+ - \delta_-$ and 
 \eqn{XiSpecial}{
  \xi=1-\beta_{+-} = -{\rm B}_{+-} =
   {y_+ y_- \over 4\pi G m_+ m_-} {k^2 \over k^2 + a^2/r_s^2}
    \,,
 }
which grows as $t^{(-1+\sqrt{1+24\xi})/6}$: faster than the adiabatic mode if $\beta_{+-} < 0$, and slower if $\beta_{+-} > 0$.  

When radiation drag and matter pressure may be neglected the baryon density contrast obeys the equation
 \eqn{BaryonsTrack}{
  \ddot\delta_b + 2 H \dot\delta_b &= 4\pi G \bar\rho_m \delta_m \,.
 }
 This standard result says that the baryons tend to approach the total mass distribution.

If we do not impose the charge neutrality condition $y_+ n_+ = y_- n_-$ then adiabaticity is not preserved and the growing mode of the mass density contrast departs from the familiar $t^{2/3}$ behavior.  To see this, consider the case where $\beta_{+-} = 0$ and ignore the effects of $r_s$ and the baryons. For this choice of parameters the gravitational attraction and scalar repulsion of unlike particles just cancel: the two species of dark matter evolve with no effect on each other. In the growing mode, $\delta_+$ varies as $t^{2\gamma_+/3}$ where $\gamma_+ = (-1+\sqrt{1+24\beta_{++} f_+})/4$, with a similar expression for $\delta_-$.  Charge neutrality implies $\beta_{++} f_+ = 1$, so in this case we recover the familiar $t^{2/3}$ growth law for the growing mode of the mass contrast, independent of the initial $\delta _\pm$.  Without charge neutrality, but still with $\beta_{+-}=0$, an initially adiabatic perturbation evolves into a mass perturbation dominated by the contrast, $\delta_+$ or $\delta_-$, which has the larger $\gamma$.  If $\beta_{+-}<0$, then without charge neutrality an initially adiabatic perturbation may evolve into one in which $\delta_+$ and $\delta_-$ tend to have opposite signs.

\subsection{A spherical model for nonlinear clustering}
\label{EJECT}

The spherical model offers a simple way to illustrate some of the effects of strongly nonlinear clustering. The numerical example presented in \cite{gpOne} shows the nonlinear growth of a mass concentration out of an initially isocurvature (homogeneous mass density) perturbation with globally neutral scalar charge. Here we are interested in the nonlinear growth of charge separation. We focus on the case of two dark matter components, $(+)$ and $(-)$, with very different charge-to-mass ratios, we assume  ${\rm B}_{++}$ is of order unity (so ${\rm B}_{--}\gg 1$), and we assume the length scales are small compared to $r_s$ so B is independent of scale. We present first some simple analytic considerations within the spherical model, and then a numerical example. 

Consider a neutral dark matter halo with initially identical distributions of the two dark matter components, in static equilibrium with no angular momentum. Since the scalar charge density vanishes there is no scalar force: the structure is the same as a standard cold dark matter halo. We will suppose almost all the halo mass $M$ is in the $(+)$ dark matter component. Now imagine varying the distribution of the light component while holding the massive component fixed. The variation of the energy of the light component under variations of its characteristic radius, $R_-$, is dominated by the scalar field potential energy, $U_-$, until the separation is large enough to make the kinetic energy of the light component important. When $R_-$ is much less than the radius $R_+$ characteristic of the $(+)$ matter the potential energy of the light component is negative, because it is dominated by the self-attraction of the $(-)$ matter. When $R_-\gg R_+$ the potential energy is positive, because it is dominated by the repulsion of the $(+)$ matter that the $(-)$ matter sees as a central point-like charge. The limiting behavior thus is 
\eqn{pelimitingcases}{
U_-\sim - {GM^2{\rm B}_{++}\over R_-}\hbox{ at }R_-\ll R_+, \quad
U_-\sim + {GM^2{\rm B}_{++}\over R_-}\hbox{ at }R_-\gg R_+.
}
One sees that there has to be an extremum of the potential energy at $U_->0$. If, as we suspect, $U_-$ does not have a local minimum near the extremum, the $(-)$ matter is unstable against slipping off the halo of $(+)$ matter. 

At central mass density $\rho _c$ the characteristic time for the light component to slip off the massive component  is 
\eqn{tminus}{
t_-\sim (G\rho _cB_{++}m_+/m_-)^{-1/2}.
}
We are assuming $m_+\gg m_-$ and $B_{++}\sim 1$, so $t_-$
is much shorter than the dynamical time for the massive component, $t_+\sim [G\rho _c]^{-1/2}$. This means a perturbation to the distribution of the light component grows by a considerable factor, $\sim\exp t_+/t_-\sim\exp\sqrt{m_+/m_-}\gg 1$, before the heavy component can react. The conclusion is that, unless isocurvature fluctuations are very strongly suppressed, a well-mixed halo is a transient phenomenon: in spherical symmetry the light component will slide into a compact core, where it finds a balance with its kinetic energy, or else it will disperse to radii $\gap r_s$, carrying with it the extra binding energy of the dark halo that ends up bound by the scalar force as well as gravity.  

We turn now to a numerical illustration of a variant of this effect, in the spherically symmetric development of charge separation during the growth of a dark halo out of a small initial density fluctuation. This solution assumes the halo is much smaller than $r_s$, and it ignores the mass in baryons. The halo is neutral, but there is a small initial charge separation, with contrast comparable to the initial mass density contrast. 

We simplify notation by scaling the particle masses and charges at fixed densities of mass and charge so $y_+=y_-$. This does not affect the dynamics, but it brings the expressions for the forces on a $(+)$ particle caused by a $(+)$ particle at distance $\vec r$ and by a $(-)$ particle at the same separation to the forms
 \eqn{accn}{
m_+ \vec g_{++} = -{\vec r\over r^3}\left( Gm_+^2+ {y^2\over 4\pi }\right)\,, 
\quad
m_+ \vec g_{+-} = -{\vec r\over r^3}\left( Gm_+m_- - {y^2\over 4\pi }\right)\,.
 }
In a spherically symmetry, the inward accelerations of the $(+)$ and $(-)$ dark matter components are 
 \eqn{totalacceleration}{
 g_\pm (r)= {G\over r^2}\left[ M_\pm (r)\left( 1+\hbox{\rm B} {m_\mp\over m_\pm }\right) + M_\mp (r)(1 -  \hbox{\rm B})\right]\,,
 }
where the masses within physical radius $r$ are $M_\pm(r)$. In a further simplification we have written $\hbox{\rm B}=\hbox{\rm B}_{+-}=y^2/(4\pi G m_+m_-)$ and, as before, we are assuming the system is small compared to $r_s$. Equation \totalacceleration\ displays the two free parameters in the physics within our assumptions.  To illustrate the situation under consideration we have chosen
\eqn{physicsparameters}{
{\rm B} = 6, \qquad m_-/m_+ = 0.01.
}
This corresponds to charge-to-mass ratios $y/\sqrt{4\pi G}m_+=0.24$ and   
$y/\sqrt{4\pi G}m_-=24$.

The remaining  free parameters are in the initial conditions. The initial integrated mass contrast in the growing perturbation mode is
\eqn{initialmasscontrast}{
\delta_{m<r}={M(<r)\over \bar\rho\, V(<r)} - 1 = 
 {3\over 5a_x}\cos ^2\left(\pi r\over 2R\right)\, .
}
This refers to the mass within radius $r$; it is to be distinguished from the mass density contrast $\delta_m = \delta\rho (r)/\rho$. The function in the last expression vanishes at the outer radius $R$ of the system, meaning the total mass is not perturbed from a background model that we take to be Einstein-de Sitter. The function is flat at $r\simeq R$ and $r\simeq 0$, so the mass densities are close to homogeneous at the center and periphery. The larger density near the center requires that there is an intermediate zone where the mass density is smaller than the background model. This situation might approximate the formation of an isolated dark halo in a region with low mean density. 

The parameter $a_x$ in \initialmasscontrast\ is defined by the evolution of the mass density at $r\ll R$ when the scalar force vanishes (or is canceled by identical distributions of the two species of charged particles). In this case, for a mass shell close to the center, $a_x$ is the ratio of the radius of the mass shell at the moment when the shell stops expanding to the radius at the initial conditions in equation~\initialmasscontrast . In our numerical example this central expansion factor is $a_x=30$. 

The initial integrated charge density contrasts in the two dark matter components are
\eqn{scalarinitialmasscontrast}{
\delta_\pm = {3\over 5a_x}(1\mp \alpha f_\mp )\cos ^2\left(\pi r\over 2R\right),
}
where the parameters $f_\pm = m_\pm /(m_++m_-)$ are the mass fractions in the two components. The initial mass contrast is $\delta_{m<r} = f_+\delta_+ +f_-\delta_-$, consistent with \initialmasscontrast . The initial charge contrast is 
\eqn{initialcharge contrast}{
\delta_c = \delta_+ - \delta_- =  {3\alpha\over 5a_x}\cos ^2\left(\pi r\over 2R\right).
}
The parameter representing a primeval charge imbalance is $\alpha = 0.5$ in our example. This means the initial density contrast in the low mass negative component is 1.495 times the initial mass density contrast, and the initial density contrast in the positive species is 0.995 times the initial mass contrast. 

 \begin{figure} 
  \centerline{\includegraphics[width=3in]{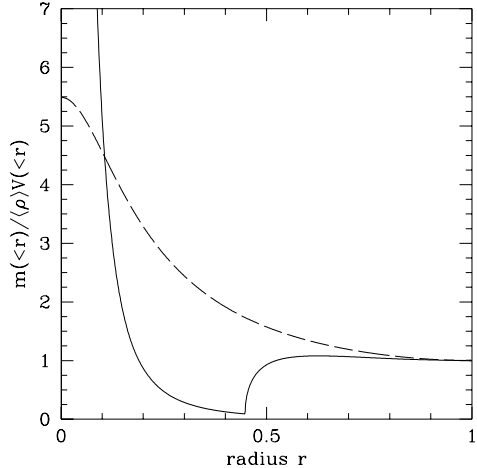}}
  \caption{A spherical model for the early development of a dark matter halo, comparing evolution in the standard cosmology --- shown as a dashed line --- and the interacting dark matter model with the parameters in equation~\physicsparameters .}\label{figD}
 \end{figure}

In figure~\ref{figD} the computed mass distributions are represented as the mean density $\bar\rho (r)$ of the mass $M(<r)$ within a centered sphere of radius $r$, as in \initialmasscontrast  . The dashed curve shows the standard model, with no scalar force, plotted at the moment when the mass near the center has stopped expanding and is about to collapse. Because mass shells are not crossing we have a simple analytic solution, which tells us that the center has stopped expanding at time $t_x = 3\pi a_x^{3/2}t_i/4$, when the central density is $\rho _{\rm inner}/\rho _{\rm outer} =(3\pi /4)^2$ times the density at edge, and where $t_i$ is the initial time in the Einstein-de Sitter model. 

The solid curve in the figure shows the effect of the scalar interaction on the mass distribution. It is plotted at the same time $t_x$ as for the dashed curve, with the same initial conditions on the mass perturbation.  We have put 1\% of the mass in the component with large charge-to-mass ratio. This light matter responds to a scalar field gradient much more rapidly than does the heavy component. The larger initial central density in the light component thus quickly drives 1\% of the mass into a tight central concentration. This central negative charge in turn pushes away the dominant mass component. In the figure there are no massive particles within the cusp in the solid curve at $r\sim 0.45R$. Massive particles at the cusp are momentarily at rest. Further out, massive particles that initially were near the center are moving outward, and the rest are still falling in to form a dilute halo around the tight concentration of 1\% of the mass. This parallels the behavior suggested by our discussion of \pelimitingcases .

If we had changed the sign of $\alpha$ in \scalarinitialmasscontrast , to place an initial positive charge density near the center, the light particles would have rapidly moved away, leaving a positively charged dark halo that is more tightly bound than in the standard model because of the scalar addition to the gravitational attraction. Again, this parallels our discussion of \pelimitingcases .

We can conclude that, for the parameters under consideration here, very small primeval isocurvature fluctuations would ensure that the charges in a dark matter halo are always separated. Under spherical symmetry the separation is to a light compact core and dispersed mass or else to a near normal dark halo and a dispersed light component. A more realistic analysis would allow for separation in the manner of a squeezed watermellon seed, but the spherical model might be a useful approximation to the first generation of halos. If the initial charge imbalance in this generation were systematic, due to a small mean excess of charge in the light component, the spherical model would suggest there is some suppression of formation of the first generation of low mass dark halos. We comment on the possible observational significance in section~\ref{DISCUSSION}. 
 
\subsection{The Void Phenomenon}
\label{VOIDS}

We consider here a possible relation between the situation we have been discussing, where most of the mass is in one of the dark matter components, and the observation that the nearby voids contain strikingly few galaxies of any  kind. The observational situation is reviewed in \cite{voids}. It may be compared to the distribution of dark matter  halos in the numerical N-body simulation shown in figure~1 in \cite{mathiswhite}. In the simulation there are relatively large dark matter halos that might be suitable homes for the normal galaxies --- with masses comparable to the Milky Way --- that contain most of the visible stars. These large halos appear in concentrations that resemble the observed clustering of normal galaxies. But the simulation shows that in the voids between the concentrations of large halos there are many relatively small halos that would seem to be suitable homes for dwarf and irregular galaxies. In the real world there are many more dwarf and irregular galaxies than normal ones, but the evidence is that they avoid the voids defined by the normal galaxies. This may not be the case in the $\Lambda$CDM cosmology: the simulations predict the presence of numerous potential homes for dwarfs in voids. Is there really a significant difference between the observed and predicted relative distributions of dwarf and normal galaxies? If so, might the scalar force help explain the difference? 

The first question is put to the test in \cite{mathiswhite}, who use the nearest neighbor statistic that is applied to the observations in \cite{voids}. The conclusion in \cite{mathiswhite} is that the predicted distribution shows little sensitivity to the halo mass, consistent with what is observed. The problem with this conclusion is that the typical distances in the simulation are an order of magnitude larger than in the observations, as one sees by comparing figure~6 in \cite{mathiswhite} to figures~4 to~6 in \cite{voids}. Thus the meaning of the comparison is not clear. We conclude therefore that there is no established quantitative demonstration of an inconsistency between theory and observation, but there are distinct visual indications of a problem. 

If there is a problem there may be a resolution within the interacting dark matter model. Let us return to the linear perturbation theory discussed in section~5.1, and consider the case $m_+\gg m_-$ and ${\rm B}_{++}\sim 1$. When the mass in the light component may be neglected, and we can also ignore the mass in baryons, the perturbation equations \FullPertEqn\ and \betaDef\ become
\eqn{limitingpt}{
&{\partial ^2\delta _+\over\partial t^2} +2{\dot a\over a}{\partial\delta _+\over\partial t}
= 4\pi G\bar\rho 
\left[\delta _+ + \left(\delta _+-\delta _- \right){\rm B}_{++}\right] \cr
&{\partial ^2\delta _-\over\partial t^2} +2{\dot a\over a}{\partial\delta _-\over\partial t}
= 4\pi G\bar\rho  
\left[\delta _+ + 
\left(\delta _- - \delta _+\right){\rm B}_{++}{m_+\over m_-} \right]\,.
}
The second equation says that the characteristic time for the evolution of $\delta _-$ is on the order of $(G\bar\rho {\rm B}_{++}m_+/m_-)^{-1/2}$, as in \tminus . Since we are assuming $m_+\gg m_-$ and ${\rm B}_{++}\sim 1$ this is much shorter than the charactristic times for the growth of the mass contrast and the expansion of the universe, both of which are on the order of $\sim (G\bar\rho)^{-1/2} $.

Under the conditions we are considering we are led to the following picture. On a time scale much shorter than the Hubble time the light component settles to minimize the sum of its potential and kinetic energy in the given distribution of the massive component. As the universe expands the kinetic energy in the light component is redshifted away, and this component tends to settle more deeply into its potential energy. The deepest minima of potential energy of the light component are in the regions of low mass density, the developing voids, where the scalar charge density of the massive component is lowest. The massive component responds to the clustering of the light component on a time scale comparable to the Hubble time.  The concentration of the light component in the protovoids promotes evacuation of dark halos from the voids. In the $\Lambda$CDM cosmology dark halos tend to leave low density regions, as part of the general growth of clustering. The problem is that the tendency seems to be weaker than wanted. The scalar force would enhance this tendency, and so perhaps resolve the void problem.  

Baryons bound to dark halos in protovoids would be encouraged to leave with the massive dark matter.  The only difference from $\Lambda$CDM in the behavior of baryons not bound to dark halos arises from the lower mass density in dark matter in voids predicted by our model --- a small effect.  It may encouraging therefore that low surface density HI clouds are observed in nearby voids (\cite{pentonetal}).

\section{Discussion}
\label{DISCUSSION}

There is little doubt that a number of constructions exist in supersymmetric field theory and string theory which give rise to scalar forces screened by light particles, as in \Particleaction.  Incorporating non-relativistic dark matter particles coupled appropriate to the scalar seems unlikely to pose a difficulty, given that wrapped branes have roughly the right properties.  Furthermore, it seems likely that the pitfall of linear terms in the Kahler potential can be avoided.  What is clearly more difficult is to build such a construction into a compactification which also includes the Standard Model and which stabilizes those moduli which couple directly to the visible sector.  Most difficult of all is to see in detail how scalar forces with sufficient range and appropriate couplings to affect large scale structure formation survive supersymmetry breaking.

Notwithstanding these challenges, it appears that string theory provides at least the right ingredients for models that would include interesting scalar forces in the dark sector.  Thus we propose a refinement of the moduli-fixing program in string theory: rather than requiring the elimination of all moduli, the objective should be to stabilize precisely those which couple to the visible sector in such a way that the absence of adequate stabilization would violate experimental constraints.  Then one should ask whether other moduli could be left unfixed by vacuum effects --- only to be stabilized by the presence of some number density of relativistic and non-relativistic particles, possibly leading to an interesting variant on the standard cold dark matter cosmological model.  If future observations decisively favor models of this type over $\Lambda$CDM, then astronomy may become a window into string physics. 

We have not provided an account of initial conditions for the variant of the cold dark matter model under discussion here.  More specifically, we have not described a mechanism which will produce the heavy dark matter particles or the relativistic screening particles in the correct abundances, or set the typical energy $\epsilon_s$ of the screening particles in the rather high range needed if $y_s$ is of order unity.  The primordial perturbation spectrum is less of a worry, since what we need is a predominantly adiabatic spectrum plus a slight isocurvature component, which can easily arise from inflation. And the most immediate issue, we believe, is the possible observational consequences or indications of a scalar force in the dark sector. The discussion in section~\ref{GROWTH} assumes most of the mass is in one of the dark matter components. We have remarked on two possible applications, to the development of small-scale structure before substantial separation of the light and heavy dark matter components, and to the development of voids as the two components become well separated. 

The illustration in figure~\ref{figD} of the possible effect of the scalar force on the formation of the first generations of dark matter halos assumes the presence of screening particles, which allows the postulate that the mean scalar charge density differs from zero, with an accompanying screening length, $r_s$, on the scalar force (as discussed in section~3). The Jeans length for the baryonic matter after decoupling, $\lambda\sim 10$~kpc, is within the bounds discussed in section~3, so we are allowed to consider observationally interesting values of $r_s$. Our illustration assumes the light component has the larger mean charge density. It suggests that this produces early formation of compact concentrations of the light component and early suppression of formation of massive dark matter halos. Our illustration also assumes spherical symmetry, however, so these indications certainly will have to be checked by numerical simulations before we may consider applying them to the apparent excess production of low mass dark halos in the $\Lambda$CDM cosmology (\cite{Klypinetal}, \cite{Mooreetal}), or to the developing constraints  from strong lensing of quasars on low mass halos in the halos of massive galaxies and in the field (as discussed in \cite{Metcalf} and references therein). 

It should be noted that the astrophysical community has not been very concerned about the issue of dark halos in voids, because the relation between dark halos and the baryons that illuminate them is difficult to predict. Many dwarf galaxies are observed in regions where the ambient density is close to the cosmic mean, however, and it is reasonable therefore to ask why so few extreme dwarf galaxies are observed in nearby voids. The proposal here is that the scalar force has pushed the massive dark matter out of the voids. Since this scenario assumes near charge neutrality and adiabatic initial conditions the scalar force has little effect on the linear evolution of the large-scale structure probed by the measurements of the anisotropy of the CBR. The effect of the scalar force on smaller scales, which is represented by the parameter ${\rm B}_{++}$, must not be too strong, because there are normal-looking galaxies in low density regions. Also, the scalar force must not upset the evidence that the Ly$\alpha$ forest, which would react to the scalar force only through its gravitational interaction with the dark matter, is a good tracer of the dark matter power spectrum (as discussed in \cite{Lyalphaforest} and references therein). It will be interesting to see numerical explorations of these effects, under the assumption that $r_s\gap 10$~Mpc.  

We have not discussed here a different parameter choice, in which the oppositely charged dark matter components have near equal particle masses. We remarked in \cite{gpOne} that if $m_-\sim m_+$ then a close pair of galaxies with dark matter halos of opposite charge could be only loosely bound, the gravitational and scalar forces nearly canceling. This with suppression of the exchange of dark matter particles between oppositely charged halos would reduce stellar dynamical drag, increasing the lifetimes of oppositely charged binary galaxies. Such an effect has some apparent observational support in the abundance of binary galaxies. In the absence of the screening particles there would be no preferred separation of binary galaxies; the distribution of separations would be set by the process of their production. With the introduction of the screening particles one can choose parameters so a pair of oppositely charged dark halos finds an equilibrium separation at $r\sim r_s$. There is no evidence of this preferred separation  in the galaxy two-point correlation function,  but a closer look at the observations might be worthwhile.

\section*{Acknowledgements}

We thank N.~Dalal, G.~Farrar, B.~Metcalf, C.~Vafa, and E.~Witten for discussions.  The work of SSG~was supported in part by the Department of Energy under Grant No.\ DE-FG02-91ER40671, and by the Sloan Foundation.

\end{document}